\documentclass[smallextended, final]{svjour3}
\usepackage{amsmath}
\usepackage[margin=1in]{geometry}
\usepackage{graphicx}
\usepackage[utf8]{inputenc}
\usepackage{soul}
\usepackage{xcolor}
\usepackage[backend=biber, style=numeric, citestyle=numeric-comp, sorting=none]{biblatex}
\usepackage{tabularx}
\usepackage{hyperref}
\graphicspath{Figures}

\title{Instance Segmentation for Direct Measurements of Satellites in Metal Powders and Automated Microstructural Characterization from Image Data}
\author{Ryan Cohn \and Iver Anderson \and Tim Prost \and Jordan Tiarks \and Emma White \and Elizabeth Holm}
\institute{ Ryan Cohn \and Elizabeth Holm \at Materials Science and Engineering, Carnegie Mellon University, 5000 Forbes Ave., Pittsburgh, PA 15213, USA
\and
Iver Anderson \and Tim Prost \and Jordan Tiarks \and Emma White \at
Materials Science, Ames Laboratory, 311 Iowa State University, Ames, IA 50011}
\date{\today}

\begin{document}
\titlerunning{Instance Segmentation for Automated Analysis of Image Data in Materials}
\authorrunning{Ryan Cohn et al.}

\maketitle
\begin{abstract}
We propose instance segmentation as a useful tool for image analysis in materials science. Instance segmentation is an advanced technique in computer vision which generates individual segmentation masks for every object of interest that is recognized in an image. Using an out-of-the-box implementation of Mask R-CNN, instance segmentation is applied to images of metal powder particles produced through gas atomization. Leveraging transfer learning allows for the analysis to be conducted with a very small training set of labeled images.  As well as providing another method for measuring the particle size distribution, we demonstrate the first direct measurements of the satellite content in powder samples.  After analyzing the results for the labeled data dataset, the trained model was used to generate measurements for a much larger set of unlabeled images. The resulting particle size measurements showed reasonable agreement with laser scattering measurements. The satellite measurements were self-consistent and showed good agreement with the expected trends for different samples. Finally, we provide a small case study showing how instance segmentation can be used to measure spheroidite content in the UltraHigh Carbon Steel Database, demonstrating the flexibility of the technique.

\end{abstract}
\section{Introduction}
\label{intro-cv}

Materials characterization and quality control rely on the analysis of microscopy images and other visual data. Manual analysis of images is a labor-intensive process and subject to human judgement. Because of this, there is growing interest in using automated computer vision techniques to analyze image data in materials science. Recent research has demonstrated how several types of computer vision methods can be used for a wide variety of applications, including identifying defects on materials \cite{Song2013, Kitahara2018, Li2018} and powder beds \cite{Scime2018, TanPhuc2019}, characterizing powder samples \cite{DeCost2017c}, segmentation of  microstructural features of interest \cite{DeCost2015Segment, Chen2018} instrument readings \cite{Stan2020}, and more \cite {Kusche2019, Ram2017, Ziletti2018, Campbell2018}. The methods in most of these studies can be categorized as classification \cite{Rawat2017}, in which a label is assigned to an image; semantic segmentation \cite{Taghanaki2019}, in which a label is assigned to each pixel in an image; or detection \cite{Zhao2019}, which indicates the class, size, and position of each instance of every object that is recognized in an image. 

Recently, researchers in computer vision have made significant advancements in the field of instance segmentation. Instance segmentation is an advanced technique in computer vision that extends object detection to include a segmentation map for each object that is recognized in an image. This provides detailed information on the number of objects in an image, as well as the position, size, and shape of each object. With the release of the Microsoft Common Objects in Context (COCO) dataset \cite{Lin2015}, which contains 328,000 labeled images with 2.5 million labeled instances, instance segmentation has become a big area of focus in the field of computer vision. 

Current approaches to instance segmentation all rely on deep learning and convolutional neural networks. Mask R-CNN \cite{He2017}, introduced by Facebook AI Research in 2017, is still one of the most popular network architectures used for the task of instance segmentation. Mask R-CNN extends Faster R-CNN \cite{Ren2015}, a network with good performance on object detection tasks,  with additional convolution layers for predicting individual segmentation masks for each instance. At the time of its release, Mask R-CNN achieved the highest score on the COCO instance segmentation challenge and allowed for near real-time mask proposals. Currently it is still recognized as a standard approach to instance segmentation and serves as a benchmark for comparing the performance of new network architectures. 

Despite being a powerful tool for automating image analysis, instance segmentation has not yet been widely applied for applications in materials science.  In this paper we provide a case study using instance segmentation to improve powder characterization with potential applications in additive manufacturing (AM.)  In powder-bed fusion AM, the properties of the feedstock powder influence the quality of the parts produced \cite{Anderson2018}. However, current methods of characterizing powder size and rheology aren't always sufficient to predict the quality of parts after a build \cite{Clayton2015}. For example it is established that satellite formation on powder particles influences flowability \cite{Sun2017}, but it is not currently possible to experimentally measure satellites on metal powders. We apply instance segmentation to scanning electron microscope (SEM) images of metal powders to generate the first direct measurements of powder satellites. We then extend this approach to show how instance segmentation can be used to measure the spheroidite content in steel microstructures, demonstrating the flexibility of the technique to apply to a wide variety of applications in materials science.
 
\section{Methods}
\subsection{Data Collection and Labeling}

\begin{figure}
    \centering
    \includegraphics[scale=0.125]{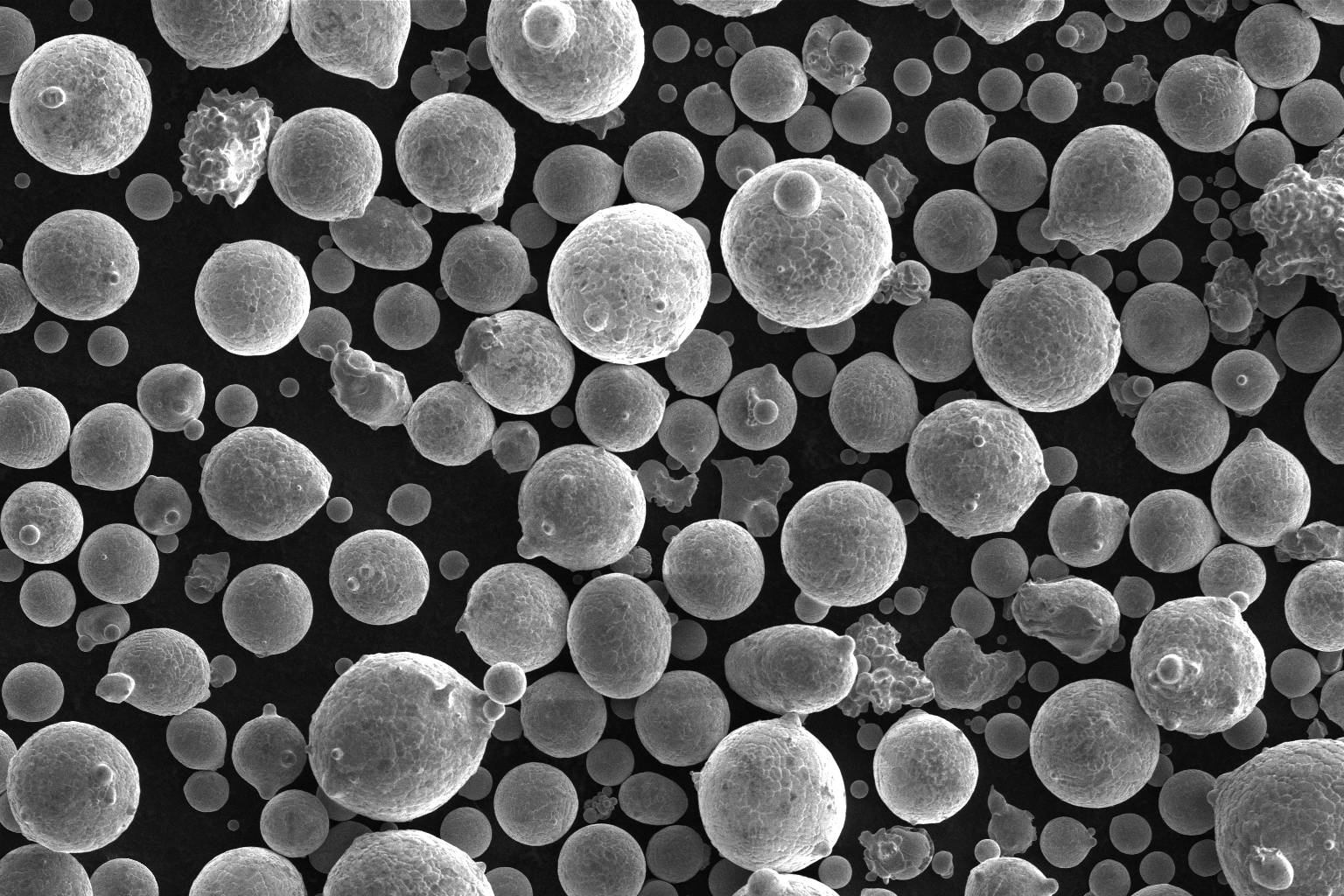}
    \caption{Sample image of powder particles. Satellites are visible on several particles.}
    \label{fig:sample_image}
\end{figure}

\begin{figure}
    \centering
    \includegraphics[scale=0.15]{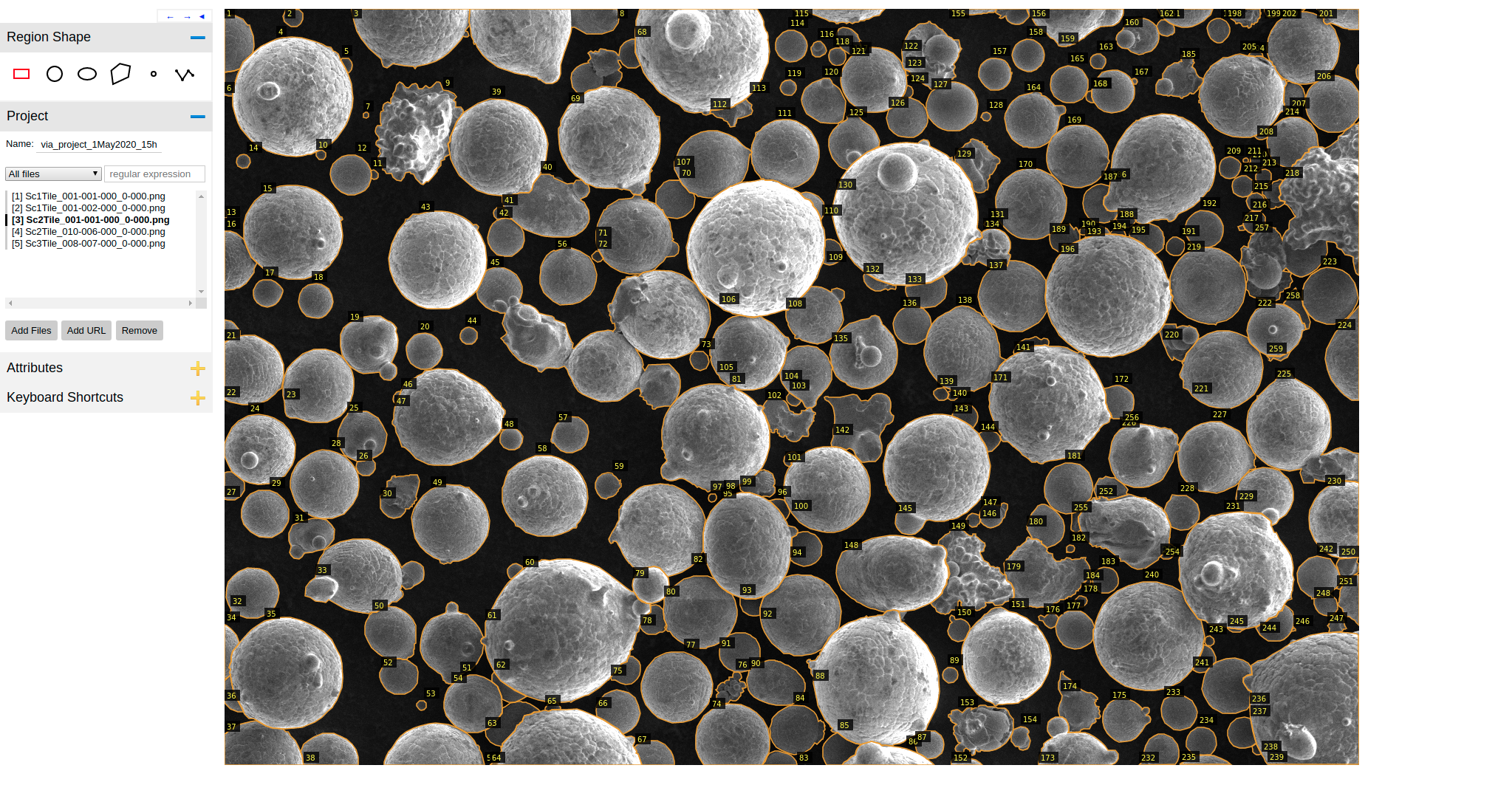}
    \caption{Screenshot from VGG Image Annotator software with labels for powder particles.}
    \label{fig:sample_annotations}
\end{figure}

Data for this study consisted of SEM images of a gas-atomized nickel superalloy powder. A sample image is shown in Figure \ref{fig:sample_image}. The VGG Image Annotator (VIA) \cite{DuttaAbhishekandZisserman2019} was used to label images for training and evaluation.  The mask for each instance was approximated by drawing a polygon around each individual powder particle or satellite. A sample screenshot of the program with annotations for powder particles in the same image is shown in Figure \ref{fig:sample_annotations}.

The bounding boxes were derived from the polygons by taking the highest and lowest X and Y coordinates from each polygon. Five images were annotated with labels for both powder particles and satellites. An additional 5 images were annotated with only labels for satellites to account for there being fewer satellites per image than powder particles. In total there were 1360 labeled powder particle instances and 1029 labeled satellite instances. After labeling, each dataset was divided into subsets for five-fold cross validation. For the powder particle dataset, four images were used to train the model and the remaining image was used for evaluation. For the satellite dataset, eight images were used for training and the remaining two were used for evaluation. In both datasets, each image was used in four of the training subsets and one of the validation subsets.

\subsection{Model Training}
\label{sec:MethodModelTrain}

Training is the process of adjusting the model parameters to minimize the loss function, which quantifies errors in model predictions. Mask-RCNN utilizes a multi-task loss function \cite{He2017} that incorporates losses from predictions in class labels, bounding box coordinates, and binary segmentation masks for each instance, shown in equation \ref{eqn:loss}.

\begin{equation}
\label{eqn:loss}
L = L_{cls} + L_{box} + L_{mask} 
\end{equation}

The first component of the loss equation, \(L_{cls} \), measures the error in the predicted class label for each instance \cite{Girshick2015}.  The class prediction branch of Mask R-CNN uses a softmax layer to output the final class predictions for each instance. For instance \(i\), the class prediction is a vector denoted \(\vec{p^i}\). Each element \(\vec{p^i_j}\) exists on the interval (0,1) and is interpreted as the predicted probability that instance \(i\) belongs to class \(j\). If the true class of an instance \(i\) is \(u\), then \(L_{cls}\) is given by the log loss function, shown in equation \ref{eqn:Lclass}:
\begin{equation}
\label{eqn:Lclass}
L_{cls}(\vec{p^i},u) = -log\hspace{2pt}\vec{p^i_u}
\end{equation}

The second term in the loss function measures differences between the predicted and true bounding boxes for each instance. The ground truth for the bounding box for an instance of class \(u\) is given by the vector \(\vec{v}=(v_x, v_y, v_w, v_h)\), where the four indices indicate the x and y coordinates of the center of the box, the width of the box, and height of the box respectively. Detailed information about the format of the bounding boxes is given in \cite{Girshick2013}.  The predicted bounding box is denoted \(\vec{t}\) and has the same form as \(\vec{v}\).  \(L_{box}\) is given in equation \ref{eqn:Lbox}:

\begin{equation}
\label{eqn:Lbox}
L_{box} = \sum_{i\in{x,y,w,h}} \text{smooth}_{L_1}(t_i^u-v_i)
\end{equation}

 In this equation, the \(\text{smooth}_{L_1}\) loss is defined by the following equation:
\begin{equation}
    \label{eqn:smoothL1}
    \text{smooth}_{L_1}(x) = 
    \begin{cases}
    0.5x^2 & \text{if } |x| < 1 \\
    |x| - 0.5 & \text{else}
    \end{cases}
\end{equation}

This loss function accounts for both the size and position of the predicted bounding box for each instance. Note that bounding boxes are always rectangular and aligned vertically, so there is no prediction or loss associated with the shape or orientation of the box.

Finally, \(L_{mask}\) accounts for differences between the predicted and ground truth binary segmentation masks for each instance. In the mask prediction branch of Mask R-CNN, a sigmoid activation is applied to every pixel in the final feature map. This bounds the values at each pixel to the interval (0,1) and is interpreted as the probability that a given pixel is included in the proposed segmentation mask for the instance. Then, \(L_{mask}\) is given by the binary crossentropy between the predicted and ground truth masks. Let \(Y_i\) and \(\hat{P}_i\) correspond to the ground truth pixel label (0 or 1) and the predicted probabilities for pixel \(i\), respectively. For ground truth and predicted masks with \(N\) total pixels, \(L_{mask}\) is shown in equation \ref{eqn:Lmask}.

\begin{equation}
    \label{eqn:Lmask}
    L_{mask} = \frac{-1}{N}\sum_{i=1}^{N} Y_i\text{log}\hat{P}_i + (1-Y_i)\text{log}(1-\hat{P}_i)
\end{equation}

Stochastic gradient descent is used to train the model. In this process, the training data are randomly split into small batches. During each iteration of training, the losses are computed on a single batch of training data using forward propagation. Then, the gradient of each parameter in the network with respect to the loss is computed using backpropagation. Finally, the gradients are used to update each parameter in the network, and the resulting set of parameters will achieve a slightly lower loss on the same batch of data. This process is repeated for many iterations throughout the duration of training.

Detectron2 \cite{wu2019detectron2}, provided by Facebook AI Research, provides a convenient and open source implementation of Mask R-CNN in python using the PyTorch framework. The pre-trained model for Mask R-CNN with a ResNet-50 backbone + Feature Pyramid Network, trained for about 37 epochs on the COCO 2017 training dataset, was obtained from the Detectron2 Model Zoo library. Models were trained to predict masks for individual powder particles and satellites. Using separate models for each class simplified the process of data labeling as separate images could be annotated for each class. For both the powder particle and satellite datasets, 5 models were trained, one for each subset of data used in five-fold cross validation. Each model was trained for 5,000 iterations using the default stochastic gradient optimizer provided in Detectron2.

To simplify the process of model training and evaluation, we present AMPIS, an open source framework for performing instance segmentation on materials data. AMPIS provides a high-level interface to  Detectron2 and provides additional tools for data evaluation and visualization. AMPIS was written with two main objectives. The first goal is to simplify the process of performing instance segmentation for materials scientists who may not be familiar with PyTorch. The second goal is to extend Detectron2 to provide useful tools specific to analyzing materials data. AMPIS includes implementations for data analysis and visualization used in this paper, including measuring the satellite content from images of powder particles. AMPIS is available at the following link: \href{https://github.com/rccohn/AMPIS}{https://github.com/rccohn/AMPIS}.

\subsection{Model Evaluation}
\label{sec:MethodModelEval}

In the COCO challenge instance segmentation models are evaluated by precision and recall scores that are averaged across different instance classes, and only account for a maximum of 100 instances per image. These scores provide a convenient way of evaluating the performance of models on large datasets with many classes. In this study, each model is trained on a small dataset with only one instance class. Thus, we propose a slightly different set of metrics that are easier to interpret for this application.

The outputs of Mask R-CNN consist of predictions for the class labels, bounding box coordinates, and segmentation masks for each instance. In order to evaluate these predictions the predicted instances must be matched with their corresponding ground truth instances. This is done on the basis of intersection over union (IOU) score, defined in equation \ref{eqn:IOU}. For two binary segmentation masks of the same size:

\begin{equation}
    \label{eqn:IOU}
    \text{IOU}(A,B) =  \frac{A \cap B}{A \cup B}
\end{equation}

In this equation, \(A \cap B\) is the number of pixels that are shared by both masks \(A\) and \(B\) (intersection), and \(A \cup B\) is the total number of pixels that are occupied by both masks (union.) Possible IOU score ranges between 0 (no overlap between \(A\) and \(B\)) and 1 (\(A\) and \(B\) are identical to each other.) 

To determine the matching pairs of instances, the IOU score was computed for for all pairs of ground truth and predicted instances. For each ground truth instance the predicted instance with the highest IOU score was found. If the score was greater than 0.5, the pair of instances instance was considered to be a true positive match. Otherwise, the ground-truth instance was considered a false negative, which is an instance that was missed by the model. After computing the matches for all ground-truth instances, the remaining unmatched predicted instances were denoted as false positives.

Then, precision and recall, defined in equations \ref{eqn:precision} and \ref{eqn:recall} respectively, were used to evaluate the instance predictions.\begin{equation}
    \label{eqn:precision}
    \text{Precision = } \frac{\text{True Positives}}{\text{True Positives + False Positives}}
\end{equation}
\begin{equation}
    \label{eqn:recall}
        \text{Recall = } \frac{\text{True Positives}}{\text{True Positives + False Negatives}}
\end{equation} 

Detection precision answers the following question: What is the likelihood that a predicted instance matches a ground truth instance?  Detection recall answers the following question: For a given ground truth instance, how likely is it that a matching instance will be predicted?

The above measurements evaluate the number of correct instance matches but do not describe the quality of agreement between the segmentation masks. To account for this, we introduce a second set of metrics, called the segmentation precision and recall. For the segmentation precision and recall, true positives are defined as pixels that are included the in both the ground truth and predicted masks. False positives are pixels included in the predicted mask but are not included in the ground truth mask. False negatives are pixels included in the ground truth mask but are not included in the predicted mask.  Segmentation precision answers the following question: If a pixel is predicted to be included in the mask, what is the likelihood it is in the ground truth mask? Segmentation recall answers the following question: If a pixel is included in the ground truth mask, what is the likelihood it is included in the predicted mask? 

\section{Results and discussion}
\label{sec:results}
\subsection{Powder particle mask predictions}
\label{sec:particle mask predictions}

\begin{figure}
    \centering
    \includegraphics[scale=1]{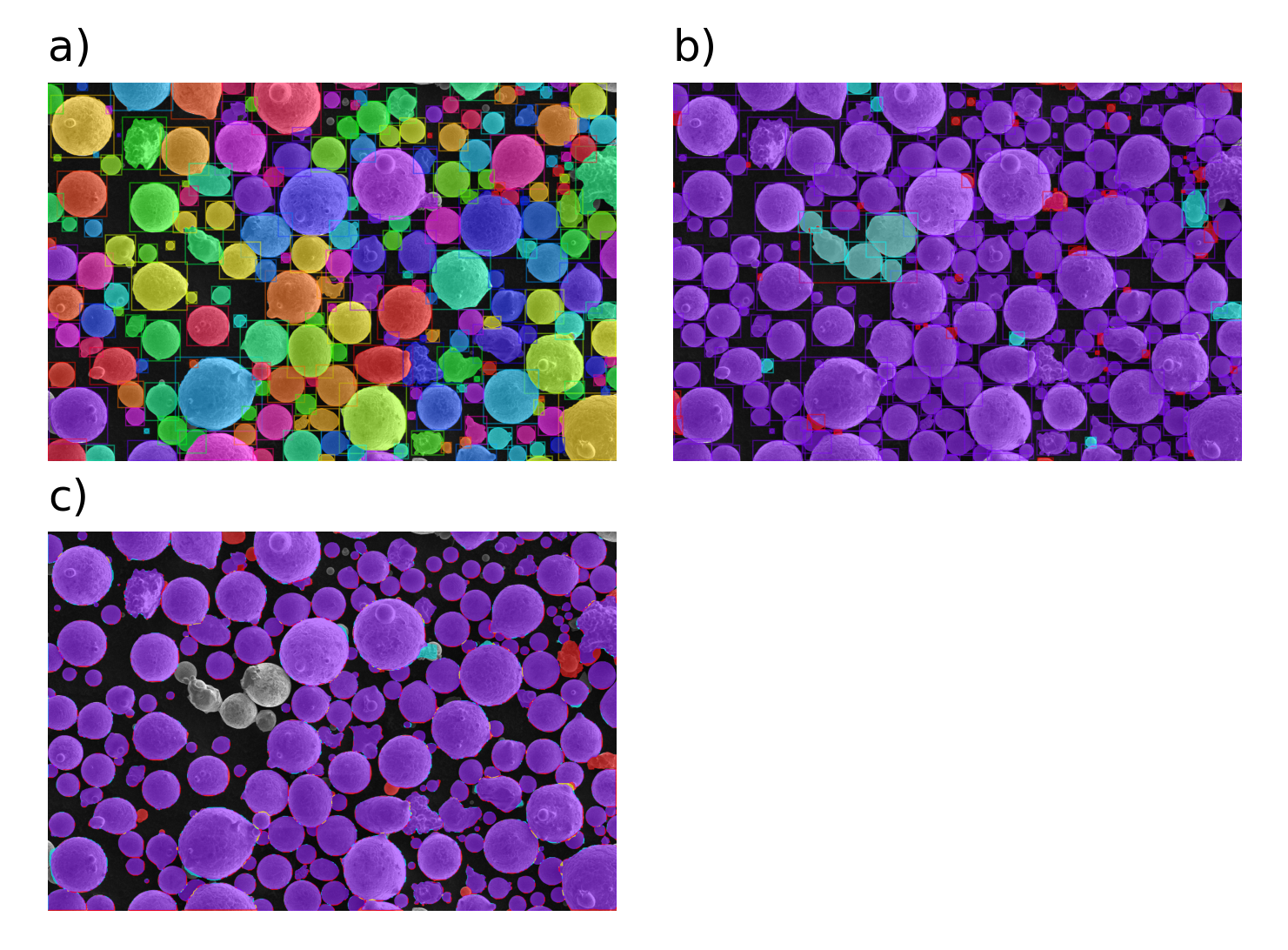}
    \caption{a) Mask and bounding box predictions overlaid on a sample validation image from cross validation. Colors are randomly assigned for visual clarity. b) Visualization of the detection performance of the predictions. True positive, false positive, and false negative instances are colored purple, blue, and red, respectively.c) Visualization of the segmentation performance of the predictions for instances that were correctly matched. True positive, false positive, and false negative pixels in each mask are shown in purple, blue, and red, respectively. Pixels that are included in multiple overlapping masks and have multiple classifications are colored yellow.}
    \label{fig:results-particle-masks}
\end{figure}

Figure \ref{fig:results-particle-masks}a shows a validation image with the mask and bounding box predictions from Mask R-CNN overlaid on the image. The colors are randomly assigned to allow for clear distinction between different instances. 
The predicted masks show very good agreement with the powder particles in the image. Figure \ref{fig:results-particle-masks}b shows the same mask predictions colored by their classification during instance matching. True positive, false positive, and false negative instances are colored purple, blue, and red, respectively. The majority of the instances are classified as true positives, confirming the strong performance of Mask R-CNN. However, there are still several false positives and false negatives in the image as well. In this experiment, false positives appear to occur as a result of the model splitting single particles into multiple smaller particles. This phenomena is common for irregularly shaped particles or large particles resulting from multiple smaller particles appearing to have fused together. Similarly, a small number of false negatives occur when the model combines particles together. This usually happens when smaller particles are directly next to large particles, and the boundary between these particles is not very clear. Finally, some false negatives occur when the model simply misses a particle. This occurs for very small, largely occluded, and irregularly shaped particle.

In most cases, false positives and false negatives occur not because the model predicts the presence of a spurious particle on the background or completely misses an existing particle. Instead, these false predictions occur as a result of a disagreement over the boundaries between particles. An example of this can clearly be seen by the group of 5 particles highlighted in blue in the middle of Figure \ref{fig:results-particle-masks}b. This collection of particles was labeled as a single particle as its components appeared to have been fused together. The model correctly recognized the presence of the particle, but split it into 5 individual particles. Therefore, even though this group of particles was recognized by Mask R-CNN, the predictions still contribute 5 false positive instances and one false negative instance to the detection scoring.  This behavior can be explained on the basis of the limited amount of training data. In each training image, the majority of particles are relatively circular and exist within a certain size range. There are only a few examples of large particles that are fused together. Therefore, during training, the model is shown many examples of mid-size, regularly shaped particles, and only a couple larger fused particles. As a result the model is more likely to recognize these regularly shaped particles than the fused particles. 

The segmentation performance for correctly matched instances is shown in Figure \ref{fig:results-particle-masks}c. For each matched pair of masks, true positive pixel predictions are shown in purple, and constitute the majority of the pixel predictions. False positive pixel predictions are shown in blue, while false negative pixel predictions are shown in red. There are a couple of larger regions containing false positive or false negative pixels. These regions can be explained by the combination or splitting of masks mentioned above. Even there is disagreement over the boundaries of particles, ground truth and predicted masks can still match with an IOU score greater than 0.5, especially if a large particle is correctly recognized and then a smaller particle is combined or split in the prediction. Therefore, the inclusion or omission of the smaller particle shows up as false positive or false negative pixels in the mask. This is why several of the blue false positive regions in Figure \ref{fig:results-particle-masks}b show up as red false negative regions in Figure \ref{fig:results-particle-masks}c and vice-versa. 

\begin{figure}
    \centering
    \includegraphics{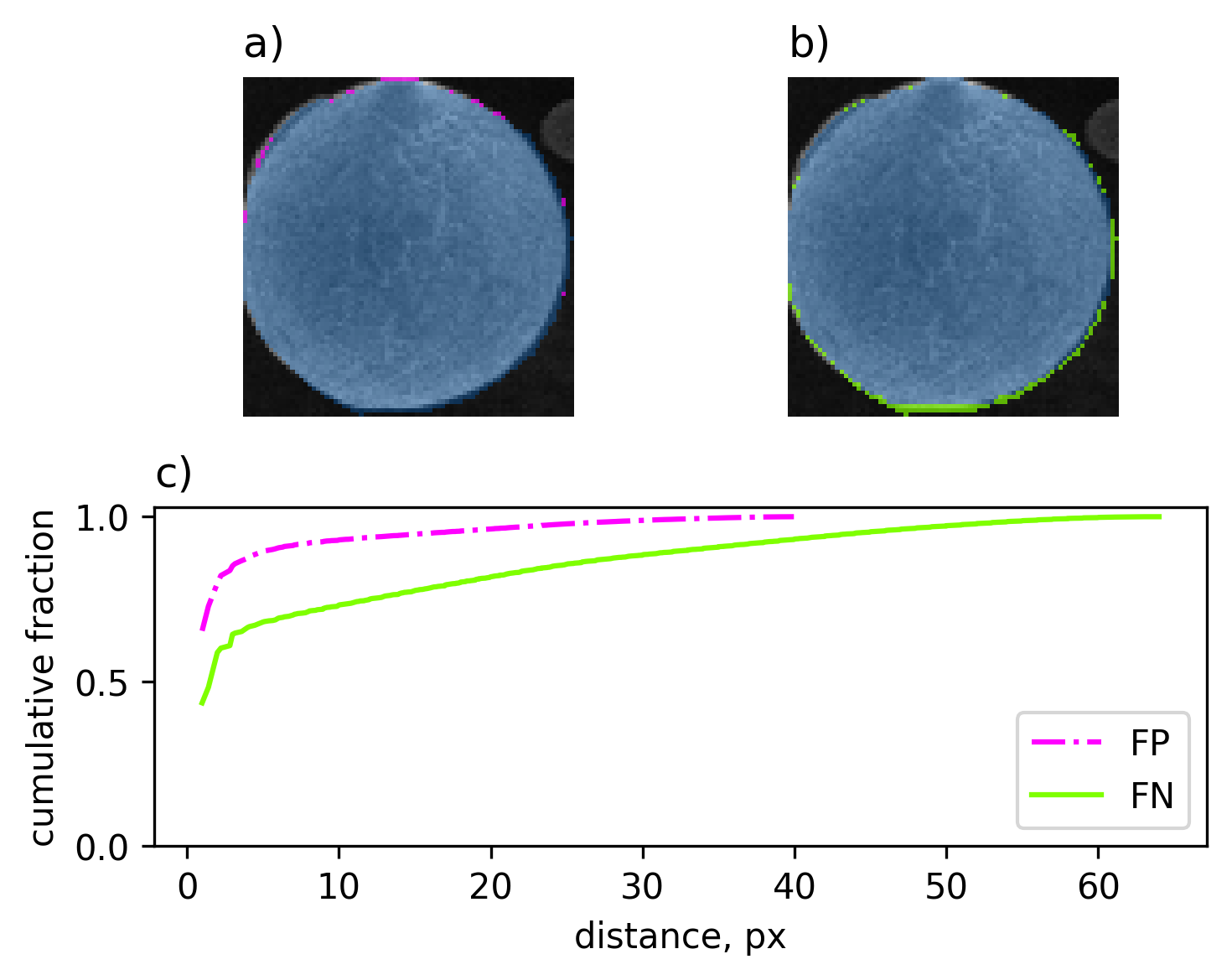}
    \caption{a) Sample image patch with ground truth mask pixels highlighted in blue. False positive pixels predicted to be included in the mask are highlighted in pink. b) The same image with the predicted mask that matched to the ground truth mask shown in a). Pixels in the predicted mask are shown in blue. False negative pixels missed in the predicted mask are highlighted in green. c) Cumulative fraction of false positive and false negative pixels in all matched instances in the image versus distance to the nearest pixel in the ground truth or predicted mask, respectively.}
    \label{fig:results-mask-dist}
\end{figure}

Additional false positive and false negative pixels appear around the edges of the particle masks. An example of this is detailed in Figure \ref{fig:results-mask-dist}. Figure \ref{fig:results-mask-dist}a shows an individual particle with the ground truth mask overlaid on the image shown in blue. False positive pixels from the predicted mask are shown in pink. Figure \ref{fig:results-mask-dist}b shows the same particle with the matched predicted instance overlaid on in blue. False negative pixels missed by the mask prediction are highlighted in green. The masks show very good agreement with only a small number of false positive and false negative pixels visible in the images.  To further quantify this agreement, the distance from each false positive pixel to the nearest true-positive pixel in the corresponding ground truth mask and the distance from each false negative pixel to the nearest true-positive pixel in the corresponding predicted mask were measured. The cumulative distribution of these distances are shown in Figure \ref{fig:results-mask-dist}c.

Since the particle labels were approximated as polygons, and the exact boundary is subjective, the predicted masks are not expected to be perfectly consistent with the labels. Even still, 65 percent of the false positive pixels are within 1 pixel of the nearest pixel in the corresponding ground truth masks, and 80 percent are within 2 pixels. Similarly, 43 percent of false negative pixels are within 1 pixel of the nearest pixel in the corresponding predicted mask, and 59 percent are within 2 pixels. Note that these measurements include the bulk regions corresponding to split or combined particles, which significantly increase these distance measurements. The measurement for false negatives is affected more as there are more of these regions for false negative pixels. The remaining predictions indicate very good agreement between the ground truth and predicted masks including the boundaries of each particle.

\begin{table}
    \centering
    \begin{tabular}{l l l l l l l l}

        cval fold 		&   0     & 1     & 2     & 3     & 4       &  avg  &  std  \\ \hline
        detection precision	&	0.944 & 0.944 & 0.921 & 0.933 & 0.946	& 0.938	& 0.010 \\
        detection recall	&	0.786 & 0.724 & 0.854 & 0.812 & 0.819	& 0.799	& 0.043 \\
        segmentation precision	&	0.987 & 0.980 & 0.976 & 0.977 & 0.986	& 0.981	& 0.004 \\
        segmentation recall	    &	0.973 & 0.966 & 0.980 & 0.974 & 0.966	& 0.972	& 0.005 \\

    \end{tabular}
    \caption{Cross-validation detection and segmentation scores for particle mask predictions. Detection scores were calculated from all instances in the validation dataset. Segmentation scores are reported as the median of all of the scores for all masks in the validation dataset. }
    \label{tab:results-particle-masks}
\end{table}

The quantitative results for all cross validation folds for the particle instance predictions are shown in Table \ref{tab:results-particle-masks}. The scores are fairly consistent across all validation folds for each category. The models achieve an average cross validation detection precision and recall of \(0.938 \pm 0.01\) and \(0.799 \pm 0.043\), respectively, and an average cross validation segmentation precision and recall of \(0.981 \pm 0.004\) and \(0.972 \pm 0.005\). Considering that labeling images is a subjective task, it is worth comparing the model results to human performance on a similar task. In \cite{Li2018}, the task of manually labeling defect loops in micrographs was repeated by five different people. It was found that humans achieved precision and recall scores of \(0.790 \pm 0.023\) and \(0.804 \pm 0.029\), respectively. Because of this variation the precision and recall of a model are not expected to ever reach perfect scores of 1 when compared to human labels.

\begin{figure}
    \centering
    \includegraphics{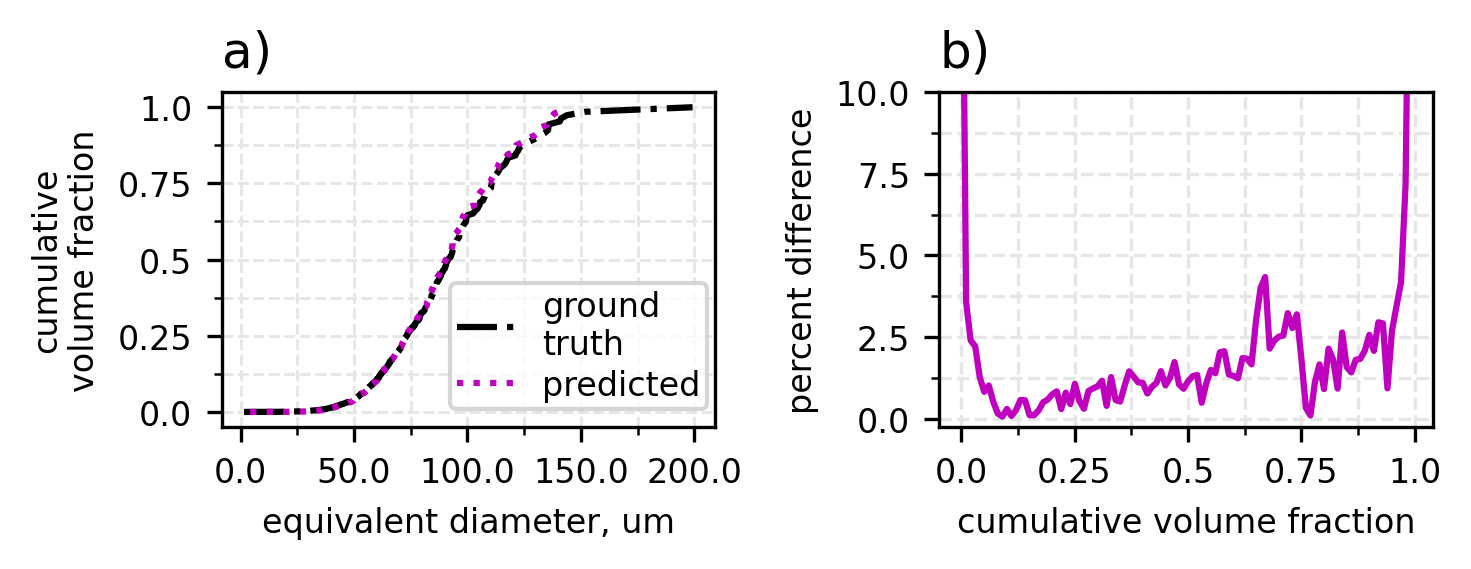}
    \caption{a) Particle size distribution determined from ground truth and predicted masks for the powder images. Particle size distribution is reported as the cumulative volume fraction versus equivalent sphere diameter. b) Cumulative volume fraction versus percent difference between distributions computed from ground-truth instances and predicted instances versus the cumulative area fraction in the distribution.}
    \label{fig:results-PSD}
\end{figure}

After generating instance predictions, the particle size distribution can be determined from the masks. To evaluate the performance of this approach, a statistically significant number of instances is required. To maximize the number of instances included in the analysis, the validation mask predictions from each cross-validation fold were combined. The predictions were compared to the ground truth labels for the same images. The resulting particle size distributions in terms of mask areas are shown in Figure \ref{fig:results-PSD}a. The distributions were interpolated between cumulative area fractions of 0.01 and 0.99 to allow for direct comparison. The percent difference between the ground truth and predicted distributions as a function of cumulative area fractions on this interval are shown in Figure \ref{fig:results-PSD}b. Between cumulative fractions of 0.01 and 0.97, the difference between the two distributions are consistently below 5\%. As the volume fraction approaches 0, the difference between the distributions rapidly increases due to the model missing the smallest particles in the labeled data. As the volume fraction approaches 1, the difference between the size distributions also rapidly increases.  In the ground truth annotations, there were two abnormally large masks  corresponding to particle diameters of 167 and 199\(\mu m\), respectively. Though the model recognized the presence of these particles, it predicted that they were actually multiple separate particles. Since the training set during cross validation did not contain any particles this large, the model was not able to recognize the presence of the abnormally large particles in the image. Thus, the difference between size distributions dramatically increases at the end of the distribution. 

\subsection{Satellite mask predictions}
\label{sec:satellite mask predictions}
\begin{figure}
    \centering
    \includegraphics[scale=0.8]{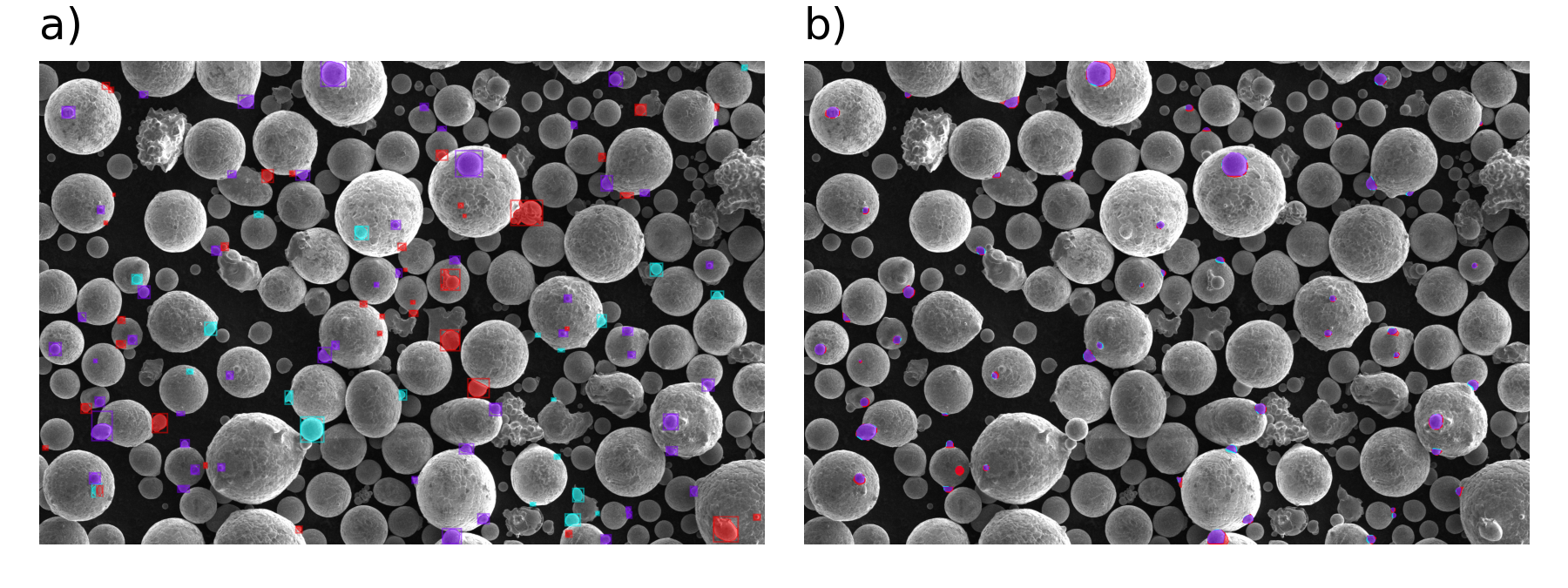}
    \caption{Satellite mask predictions. a) Visualization of the match performance of the predictions. True positive, false positive, and false negative instances are colored purple, blue, and red, respectively. b) Visualization of the mask performance of the predictions for instances that were correctly matched. True positive, false positive, and false negative pixels in each mask are shown in purple, blue, and red, respectively.}
    \label{fig:results-satellite-masks}
\end{figure}

The satellite prediction and detection performance for the same sample image are visualized in Figure \ref{fig:results-satellite-masks}a. In this figure, purple masks are true positive predictions and blue masks are false positive predictions. Red masks are false negative ground truth masks that did not match with any of the predictions. Compared to the powder particle masks, the cause of false positives and false negatives is much more straightforward. The network simply misses some labeled satellites in some cases and predicts the presence of extra satellites in other cases. The segmentation performance of matched instances is shown in Figure \ref{fig:results-satellite-masks}b. Similar to the results for powder particles, the predicted masks show very good agreement with the ground truth instances, and after matching there are disagreements between the edges of the predicted and ground truth instances.

\begin{table}
    \centering
    \begin{tabular}{l l l l l l l l}
        cval fold		&   0     & 1     & 2     & 3     & 4     & avg    & std   \\ \hline
        detection precision	&	0.732 & 0.740 & 0.720 & 0.573 & 0.693 & 0.692  & 0.061 \\
        detection recall	&	0.502 & 0.566 & 0.545 & 0.523 & 0.589 & 0.545  & 0.031 \\
        segmentation precision	&	0.909 & 0.959 & 0.948 & 0.954 & 0.886 & 0.931  & 0.029 \\
        segmentation recall     &	0.881 & 0.830 & 0.848 & 0.833 & 0.928 & 0.864  & 0.037 \\
    \end{tabular}
    \caption{Cross-validation detection and segmentation scores for satellite predictions. Detection scores were calculated from all instances in the validation dataset. Segmentation scores are reported as the median of all of the scores for all masks in the validation dataset. }
    \label{tab:results-satellite-masks}
\end{table}

The quantitative cross validation results for the satellite mask predictions are shown in Table \ref{tab:results-satellite-masks}. The models achieve an average detection precision of \(0.692 \pm 0.061\) and an average detection recall of \(0.545 \pm 0.031\). Both the precision and recall are considerably lower than the results for the powder particle predictions. Note that there is no established method for consistently counting satellites in an image. There are several particles which are clear examples of satellites, and several that clearly do not contain any. However, there are many particles with irregular bumps and features that are between these two limiting cases. For the same reason, the boundaries of satellites (i.e. the exact point where the satellite ends and the bulk particle begins) are also not clear.  During labeling the judgement of whether or not a particle contains a satellite and determining the exact boundaries of each satellite is highly subjective. Thus, neither the human nor the computer tasked with labeling the data is expected to be able to perform this task perfectly.  Instead, after training, the model will produce consistent and objective instance predictions. From looking at the mask predictions in Figure \ref{fig:results-satellite-masks}a, the predictions of satellite masks look reasonable. The model consistently labels more than half of the ground truth particles. The remaining false positive predictions, highlighted in blue, consist of small particles that touch adjacent bigger particles or bumps on the edges of particles. Though these were not not labeled as satellites, these predictions are qualitatively visually similar to many of the ground truth instances that were labeled as satellites.

\begin{figure}
    \centering
    \includegraphics{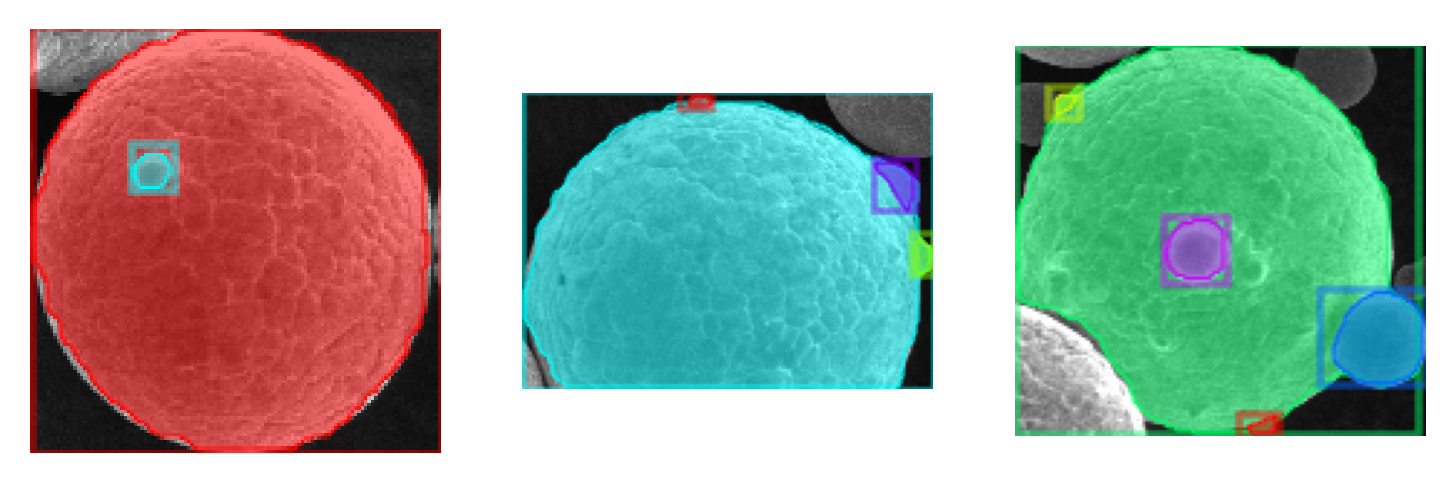}
    \caption{Sample particle masks visualized with their corresponding satellite masks after matching. Mask colors are randomly selected for visual clarity.}
    \label{fig:results-psi-match}
\end{figure}

There were 587 labeled satellites in the labeled training images, but the model predicted that there were 436 satellites, so there is about a 25\% difference between the labeled and predicted values. However, it was observed that many satellited particles contain multiple satellites. Thus, the fraction of particles that contained at least one satellite was proposed as a new metric. To match particles to satellites, the masks for powder particles and satellites for each image were overlaid. For each satellite mask, the intersection scores of each particle mask were computed. If none of the intersection scores were above 0.5, or at least half of the area of the satellite mask, the satellite was considered unmatched. Otherwise, the particle mask with the highest intersection score was considered a match for the satellite mask. Visualizations of some representative powder-satellite matches are shown in Figure \ref{fig:results-psi-match}. After computing the matches, the ratio of satellited particles is simply the number of particles that matched at least one satellite divided by the total number of particles. For the validation images in the training set, the ratio of satellited particles determined from the ground truth and predicted labels were found to be 0.232 and 0.240, respectively, so the results agree to within 3.5\% of each other. Note that smaller particles tend not to have satellites, and the model misses many of the small particles. This increases the ratio of satellited particles in the predicted masks. To remove the effect of missing small particles, the analysis was conducted after excluding all particle masks smaller than \(20 \mu m\), which accounts for more than half of the false negative ground truth instances missed by the model predictions. In this analysis, the ratio of satellited particles in the ground truth and predicted sets was found to be 0.272 and 0.255 respectively, and the results still agreed to within 7\%.

\subsection{Bulk sample measurements}

\label{ss:results-bulk}
\begin{figure}
    \centering
    \includegraphics{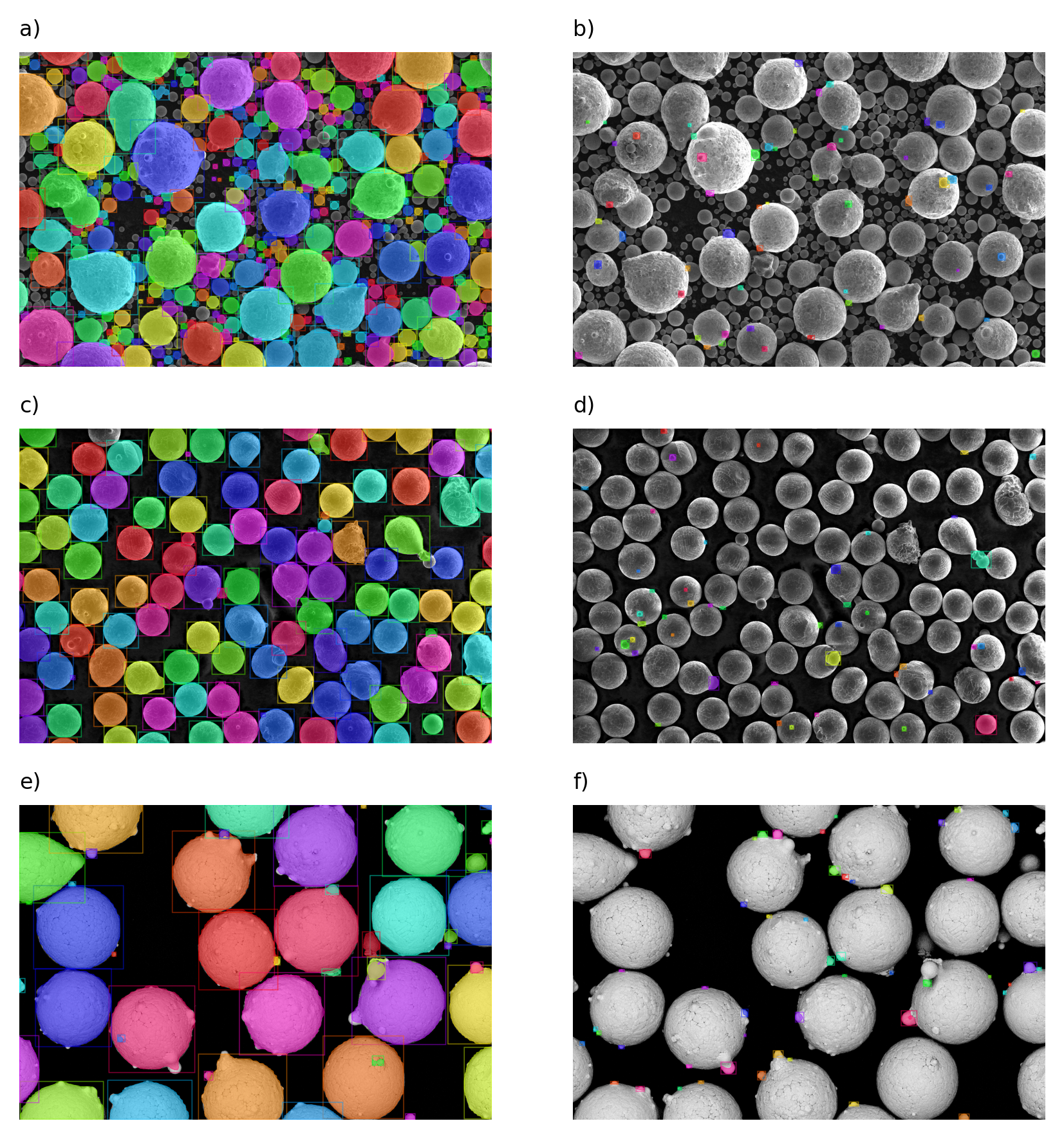}
    \caption{Sample images from model inference on larger sets of unlabeled data. a) Image from experiment one with mask predictions overlaid in color. b) Same image with satellite masks overlaid in color. c) Image from secondary electron detector in experiment two with mask predictions overlaid in color. d) Same image with satellite predictions overlaid in color. e) Image from backscatter electron detector in experiment two with mask predictions overlaid. f) Same image with satellite masks overlaid in color.}
    \label{fig:sample_inferences}
\end{figure}
\begin{figure}
    \centering
    \includegraphics{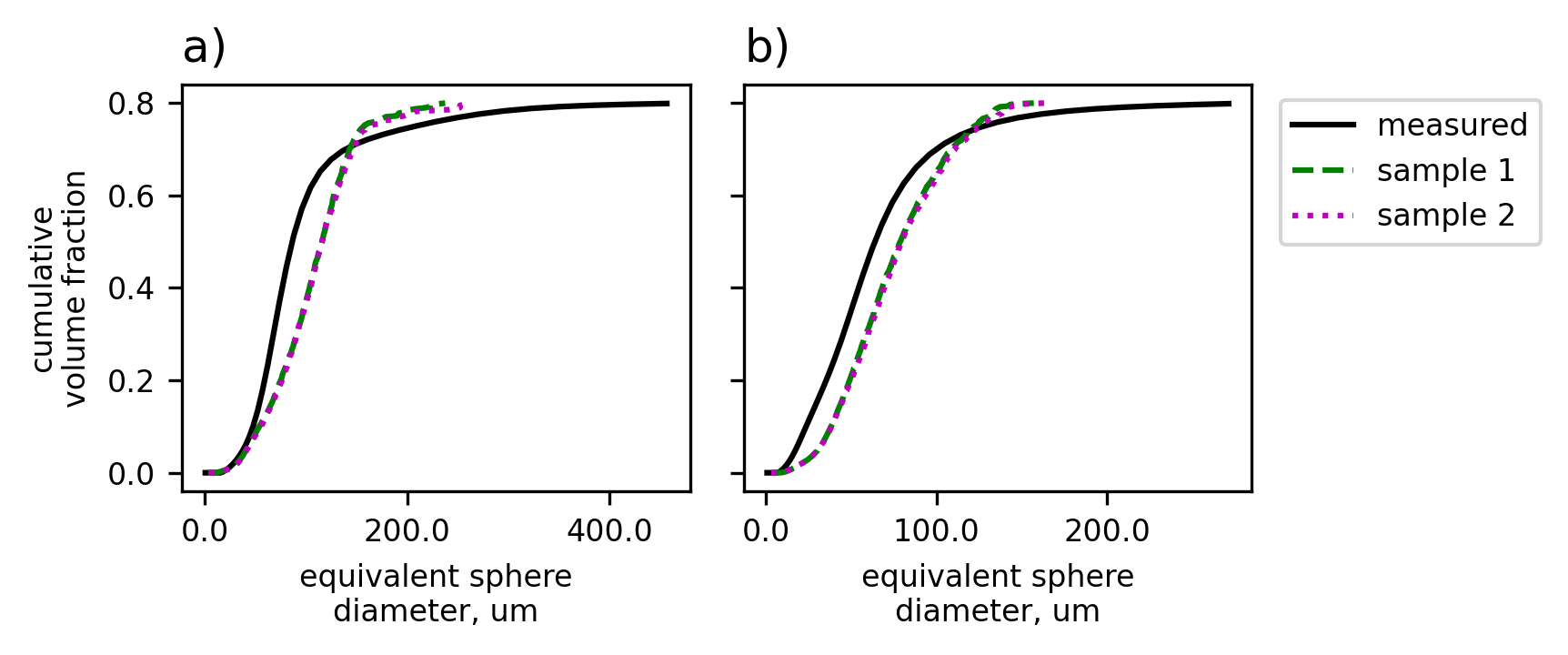}
    \caption{Particle size distributions measured from Mask R-CNN and from laser scattering for samples 2 (a) and 4 (b).}
    \label{fig:results_PSD}
\end{figure}

After characterizing the performance of Mask R-CNN on a small subset of labeled data, the model was used to generate instance predictions on two larger sets of unlabeled images. Sample images with powder and satellite masks overlaid in colors are shown in Figures \ref{fig:sample_inferences}a-f. The image in Figures \ref{fig:sample_inferences}a and b is from the same sample as the labeled training images, and was taken with the same magnification. The image has more fine particles, but are otherwise similar to those in the training set. The mask predictions for both particles and satellites show similar trends to the results for the training images discussed in sections \ref{sec:particle mask predictions} and \ref{sec:satellite mask predictions}. The image in Figures \ref{fig:sample_inferences}c and d is from a different powder sample and was taken with a magnification 20\% higher than that used in the training images. The image contains primarily mid-sized particles, and the model recognizes nearly all of them with very good performance. There are not many obviously satellited particles in the image, but the model is able to detect satellites on a couple particles. The image in Figures \ref{fig:sample_inferences}e and f are also from a set of images with 20\% higher magnification than the original training images, and are also recorded with the backscatter detector.  The image contains larger particles, most of which contain several satellites. The model appears to recognize all of the particles, but also predicts some apparent satellites to be full particles. The model also recognizes satellites on most of the particles.  Interestingly, despite changing the magnification and imaging mode, the original model predicts reasonable particle and satellite masks on these images without additional labeling and training efforts.

When generating mask predictions, the model detects particles that intersect the edges of the image. In image analysis, it is often considered good practice to remove instances that contact the edges of the image. This is because they are only partially visible and thus introduce bias into measurements like particle size. In this study, removing edge instances 
did not significantly affect the size distribution measurements, and actually increased the variation in the satellite measurements Therefore, edge particles were kept during this analysis.

The particle size distribution for two powders in the first image dataset were determined from the areas of each powder particle mask. The results were compared to the particle size distribution measured with a Microtrac Bluewave laser scattering particle size analyzer, as shown in Figure \ref{fig:results_PSD}. For both of the powder samples, the distributions show very good agreement. However, there are some systematic differences between the distributions determined from instance segmentation and the ones measured through laser scattering. 
There are systematic errors associated with different experimental methods of measuring particle size distributions, so it is not expected that distributions measured with two different methods will match exactly. Despite the differences observed on both tails of the distribution, the computer vision measurements and laser scattering measurements agree on the overall shape of the particle size distribution. 

\begin{figure}
    \centering
    \includegraphics{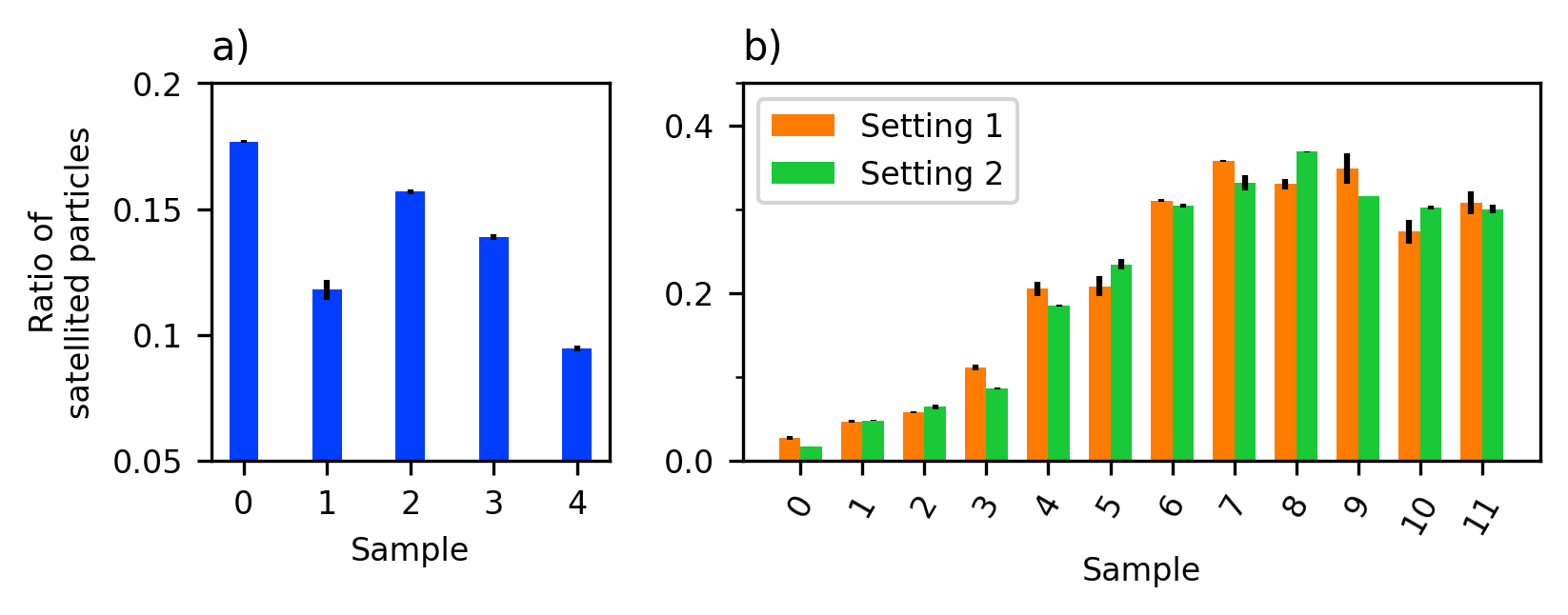}
    \caption{Ratio of satellited particles measured for each sample in two experiments. Bar heights show the average value measured from two subsets of images. Error bars show the maximum and minimum value from the subsets. a) Results for experiment 1. Each sample corresponds to a different set of atomizer conditions. b) Results for experiment 2. `Setting 1' and `setting 2' correspond to two different parameters used during atomization. The particle size increases with sample number.}
    \label{fig:results-satellite_ratio}
\end{figure}

Using the same procedure outlined in Section \ref{sec:satellite mask predictions}, the predicted particle and satellite masks were combined to determine the fraction of satellited particles in both datasets. The results for the first dataset are shown in Figure \ref{fig:results-satellite_ratio}a. The bar heights show the average value obtained from the two subsets, and the error bars show the minimum and maximum values. For each sample the results from the two subsets of images show good consistency with each other. The measurements for sample 1 agree to within 5.2\%, and the measurements for all other samples agree to within 3\%. This demonstrates the ability of the approach to generate consistent, repeatable measurements of satellites contained in powder images. For example, based on sample preparation sample 4 was expected to contain more fine particles and fewer satellites than the other samples. The measurements in this study agree with this expectation, as the fraction of satellited particles in sample 4 was 25-90 percent lower compared to the other samples. 

The second dataset contains images from two different powders produced with different atomizer settings.  Each powder was divided into 12 samples by particle size before imaging.  Once again each sample was divided into two subgroups to verify the precision of measurements and avoid double-counting particles in overlapping regions between images.  

The measured satellite content in each sample is shown in Figure \ref{fig:results-satellite_ratio}b. The difference between the two trials for each samples was less than 4\% for 20 out of 24 of the samples, indicating that the measurements are consistent. For both atomizer settings, satellite content increases with sample number, corresponding to average particle size, for the first 7 samples. This is consistent with the expected trend for satellite content. Small particles have less area and solidify faster in the atomizer column. Thus, they are much less likely to acquire satellites during atomization and are expected to have a smaller ratio of satellited particles than samples with larger particles. There is not a consistent difference in satellite content for the samples produced with the two different settings on the atomizer. These results demonstrate the potential of how instance segmentation can provide a more complete understanding of how different conditions during atomization can affect the quality of metal powders.

\subsection{Generalization to other powder images}
By far the most time-consuming task in training an instance segmentation model is manually annotating the training images. Thus, we investigate whether the trained model can be used for other particle image datasets without retraining on additional annotated images. This concept is referred to as ``transfer learning."

\begin{figure}
    \centering
    \includegraphics{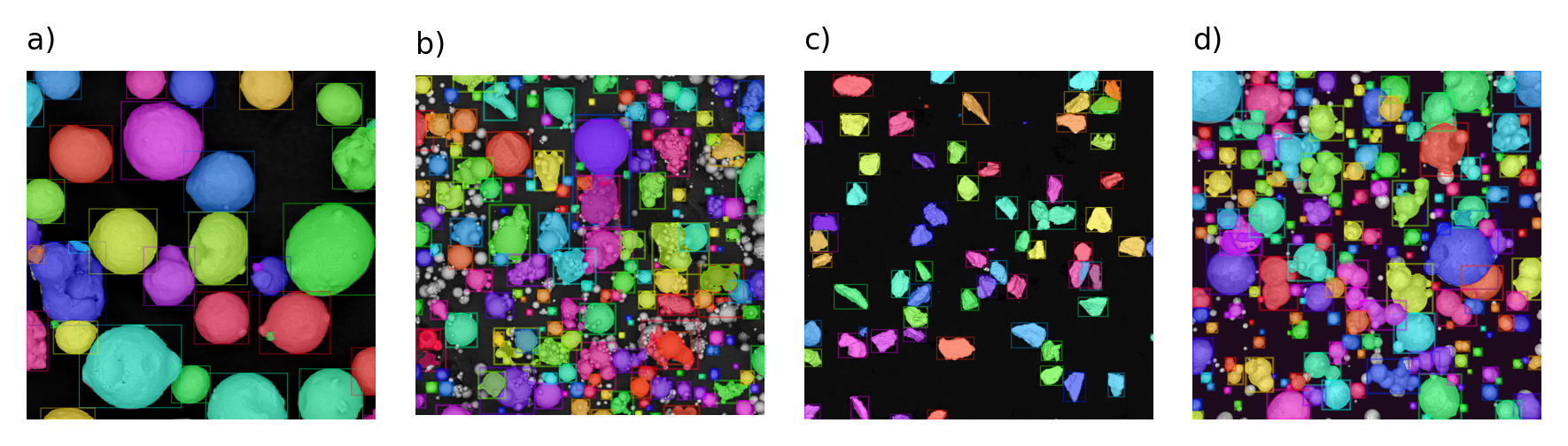}
    \caption{Model predictions on images of different samples from different experiments. Note that the original model from this study was used to generate predictions without any additional training. a) Image of commercially produced gas-atomized powder. b) Heavily-satellited Al10SiMg powder. c) Ti64 powder produced through hydride-dehydride process. d) Simulated powder particles from a synthetic dataset generated with computer rendering software \cite{DeCost2016}. 
}
    \label{fig:results-generalization}
\end{figure}

Instance segmentation of four different powders is shown in Figure \ref{fig:results-generalization}. Note that the model was not retrained on any additional labeled data before generating predictions on these images. The predictions on these images show similar trends to those described in Section \ref{sec:results}. The model recognizes most of the particles, but tends to miss some of the smaller particles and classify some of the satellites as separate particles. 

In Figure \ref{fig:results-generalization}c Ti64 powder produced through a hydride-dehydride process \cite{Narra2020} has very irregularly shaped particles. Nonetheless, the model recognizes these particles and draws reasonable particle boundaries on almost all of the particles in the image. The image in Figure \ref{fig:results-generalization}d contains synthetic powder particles generated by rendering software from \cite{DeCost2016}. The model is still able to recognize individual particles in the image, despite being a simulated micrograph. With only 4 training images, the model was able to generate useful predictions on the validation image, other images of the same samples taken with different magnification and imaging modes, images of other powder samples from different studies, and even synthetic images of simulated powders. These results demonstrate the power of Mask R-CNN to generalize to a wide variety of visual data in materials science.

We note, however, than  the satellite model did not generalize as well and tended to generate very few mask predictions for each image. This is consistent with the understanding that satellite detection is a much more subjective task and depends much more on the local visual features present on the powder particles.

\section{Instance segmentation for microstructural characterization}
\begin{figure}
    \centering
    \includegraphics{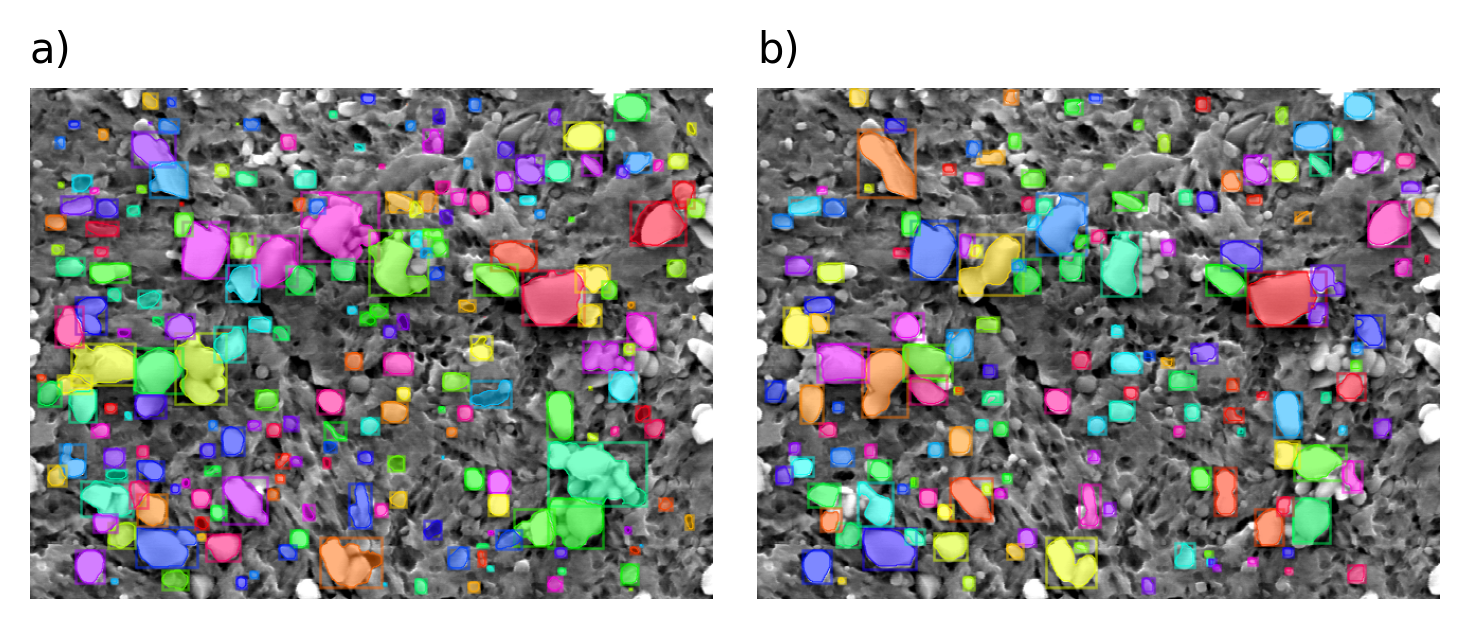}
    \caption{Sample micrograph of spheroidized steel from the validation set. a) Ground truth instance masks for spheroidite particles determined from semi-automated process described in \cite{DeCost2018uhcs}. Note that in the annotations provided in the dataset, particles that intersect the edges of the image were not included. b) Predicted instance masks from Mask R-CNN.}
    \label{fig:results-spheroidite}
\end{figure}

The above study demonstrates how instance segmentation can be used to characterize powder samples.  However, it should be noted that instance segmentation is a flexible technique that can be applied to other kinds of micrographs to automatically segment different phases, defects, or other features of interest. The UltraHigh Carbon Steel Micrograph DataBase \cite{Hecht2016, DeCost2017uhcs, DeCost2018uhcs, } provides a standard set of scanning electron microscope images of various steel microstructures, along with annotations of the constituents of each image, which can be used to test and evaluate the performance of computer vision methods. The dataset includes images of steel that contain spheroidite particles in the pearlite matrix. Mask R-CNN was trained to identify spherodiite particles on 8 images before being tested on a separate validation image. 
Figure \ref{fig:results-spheroidite}a shows the validation image with the ground-truth annotations provided in \cite{DeCost2018uhcs}; Figure \ref{fig:results-spheroidite}b shows the validation image with the predicted masks from Mask R-CNN. The model achieves a detection precision and recall of 0.70 and 0.48 on the validation image, respectively. The trends observed for detection scoring were similar to those observed for the powder particles. False positives generally result from combining or splitting neighboring spheroidite particles in a disagreement over the boundaries of the masks. False negatives occur both from the same disagreements over boundaries and also from missing smaller particles. 

In Section 3.2 of \cite{DeCost2017uhcs}, the spheroidite content in the images is characterized through semantic segmentation, achieving a cross-validation precision and recall of \(0.746\pm 0.028\) and \(0.703 \pm 0.043\), respectively. To compare the results of Mask R-CNN with the semantic segmentation methods used in this study, all of the predicted masks were combined into a single instance containing all of the spherodiite pixels.  Using this approach, Mask R-CNN achieves a precision and recall of 0.939 and 0.700, respectively. Without any parameter tuning or additional input, the model achieves a significantly improved precision while maintaining similar recall to the approach in \cite{DeCost2017uhcs}. This small study demonstrates the flexibility of Mask R-CNN in analyzing microstructural images.
\clearpage
\section{Conclusion}
Mask R-CNN was used for the task of instance segmentation on scanning electron microscope images of metal powder particles. Two separate models were trained to segment the individual powder particles and satellites in each image, respectively. Transfer learning was leveraged to train each network using only 5 images for particle instances and 10 images for satellite instances. The powder particle predictions showed good performance, achieving a cross-validation detection precision and recall of 0.938 and 0.799, respectively. False positives often occurred from splitting large particles that were thought to be fused together. False negatives mostly occurred from missing small or heavily occluded particles which did not contain a strong visual signal in the image. Satellite predictions scored lower, with a detectino precision and recall of 0.692 and 0.545, respectively, highlighting the subjective nature of identifying satellites. The models were used to characterize the particle size distribution and satellite content of larger batches of unlabeled images. The particle size distributions were comparable to experimental measurements from laser diffraction despite systematic error due to under-prediction of both very fine and very large particles. Additional labeling efforts are likely required to improve the detection performance for particles with sizes on the tail ends of the distribution. Overlaying the particle masks with satellite masks allowed the fraction of satellited particles to be directly measured for the first time. The satellite content measurements were self-consistent to within about 5\% for most samples. In both datasets used in this experiment, the relative satellite contents measured for different samples followed the expected trends. Mask R-CNN was also used to segment spheroidite particles from the UltraHigh Carbon Steel database, achieving a segmentation precision and recall of 0.939 and 0.700. The results represent a significant improvement in precision compared to a previous approach in the literature while maintaining a comparable recall.

The results from these experiments demonstrate how instance segmentation can be a useful tool for automating image data for a variety of applications in materials science and can characterize samples in ways that are not possible with other approaches. Ongoing research efforts strive to continue improving the performance of instance segmentation and continue pairing computer vision measurements with experimental results to enhance the fields of research and development as well as quality control.

\section*{Acknowledgments}

This work was supported by the National Science Foundation under grant CMMI-1826218 and by the Air Force Research Laboratory under cooperative agreement number FA8650-19-2-5209. 

\section*{Disclaimer}

The views and conclusions contained herein are those of the authors and should not be interpreted as necessarily representing the official policies or endorsements, either expressed or implied, of the Air Force Research Laboratory or the U.S. Government. 

\section*{Conflict of interest statement}
On behalf of all authors, the corresponding author states that there is no conflict of interest.

\printbibliography

\end{document}